\pdfoutput=1
\documentclass[11pt,a4paper]{article}

\usepackage[margin=2.5cm]{geometry}
\usepackage[T1]{fontenc}
\usepackage[utf8]{inputenc}
\usepackage{lmodern}
\usepackage{microtype}
\usepackage{amsmath,amssymb,amsthm}
\usepackage{graphicx}
\usepackage{booktabs}
\usepackage{array}
\usepackage{tabularx}
\usepackage{xcolor}
\usepackage{tikz}
\usetikzlibrary{arrows.meta,positioning,shapes.geometric,calc,decorations.pathreplacing,fit,backgrounds}
\usepackage{tikz-cd}
\usepackage[round,authoryear]{natbib}
\usepackage{caption}
\usepackage[colorlinks=true,linkcolor=blue!50!black,citecolor=blue!50!black,urlcolor=blue!50!black]{hyperref}
\usepackage[capitalise,nameinlink]{cleveref}

\theoremstyle{definition}
\newtheorem{definition}{Definition}
\newtheorem{example}[definition]{Example}
\newtheorem{remark}[definition]{Remark}
\theoremstyle{plain}
\newtheorem{proposition}[definition]{Proposition}

\newcommand{\type}[1]{\ensuremath{\ulcorner}\text{#1}\ensuremath{\urcorner}}
\newcommand{\cat}[1]{\mathbf{#1}}
\newcommand{\op}{\mathrm{op}}
\newcommand{\N}{\mathbb{N}}
\newcommand{\R}{\mathbb{R}}

\tikzset{
  olog/.style={draw, rectangle, align=center, font=\small, inner sep=4pt, minimum height=7mm},
  place/.style={circle, draw, thick, minimum size=7mm, inner sep=0pt, font=\small},
  trans/.style={rectangle, draw, thick, fill=black!10, minimum width=4mm, minimum height=7mm, inner sep=1pt, font=\scriptsize},
  arr/.style={-{Stealth[length=2.2mm]}, thick},
}

\title{Applied Category Theory for Genomics\\ -- An Initiative}
\author{Yanying Wu\\[4pt]
\small Centre for Neural Circuits and Behaviour, University of Oxford, UK\\
\small\texttt{yanying.wu@cncb.ox.ac.uk}}
\date{26 November 2026}

\begin{document}
\maketitle

\begin{abstract}
The genome of an organism is the totality of its DNA, and what we know about genomes is now held in large and heterogeneous data: sequences and their annotations, gene expression, regulatory interactions, three-dimensional chromatin contacts and curated ontologies of gene function. A central challenge for genomics is to integrate these data into a coherent understanding of how genomes are organised. Applied category theory (ACT) is built around two ideas that address this challenge directly: \emph{compositionality}, the description of a complex system by how it is assembled from simpler parts, and \emph{functorial semantics}, the separation of the structure of a system (its syntax) from its interpretation (its semantics) by a structure-preserving map. In this paper we argue that genomics and ACT are a natural match, and we take first steps towards a \emph{categorical genomics}. After reviewing the development of both fields, we identify three features of genomic data (heterogeneity, hierarchy and interconnection) and the categorical tools suited to each. We then present three bridges between the fields, drawn from our earlier work: a categorical compositional distributional (DisCoCat) model in which the function of a protein is computed from its domains; open Petri nets, with which gene regulatory networks are assembled from modules glued along shared genes; and Gene ologs, in which the Gene Ontology and its annotations become categories and their instances. We show that the three bridges share a single pattern: a category of syntax, an operation of composition, and a semantics functor. Finally, we construct a positional category of genes from a reference genome. Genes on a chromosome arm form an interval order, whose opposite is exactly the order seen from the complementary strand, and the familiar order by genomic coordinate is a linear extension of it. We indicate how synteny, chromosomal inversions and nested chromatin domains are described by functors, and how the four views can be connected through a common database schema. The paper is written to be readable by both biologists and mathematicians; no prior knowledge of either field is assumed.
\end{abstract}

\tableofcontents

\section{Introduction}\label{sec:intro}

DNA is the hereditary material of almost all living organisms. The complete set of DNA sequences of an organism is its genome, the store of information from which the organism, be it a human or a fly, is built and maintained \citep{Mawer2003}. Genomics, the study of the content, organisation and function of genomes, has therefore been central to biology since its beginnings.

The twentieth century saw three milestones on the way to modern genomics: the discovery of Mendel's laws of inheritance \citep{Mendel1866}, the double-helix structure of DNA \citep{Watson1953}, and the first drafts of the human genome sequence \citep{Lander2001,Venter2001}. Since then, high-throughput sequencing and bioinformatics have produced an enormous and still growing amount of data about genomes, from complete reference sequences to single-cell measurements of gene expression and maps of chromatin folding. A major challenge for genomics is now to integrate this information and to obtain a unified view of how a genome is organised and how it works \citep{Hawkins2010}. Many computational methods have been brought to bear on this challenge, most recently deep learning and large language models trained on DNA and protein sequences \citep{Zou2019,Lin2023,Nguyen2024}. These methods are powerful at prediction. They are less suited to making explicit \emph{how} the parts of a biological system are put together, and how knowledge obtained at one level or from one kind of data relates to knowledge at another.

Category theory is a branch of mathematics that is concerned precisely with structure and composition. It was introduced by \citet{EilenbergMacLane1945} to make precise the notion of a ``natural'' transformation in algebraic topology, and it soon became a common language for many areas of mathematics \citep{Awodey2010,Riehl2016}. Over the past three decades it has been applied with increasing success outside mathematics, first in computer science, physics and linguistics \citep{BaezStay2011,CoeckeKissinger2017,Coecke2010}, and more recently in chemistry, systems biology, databases and many other areas \citep{Spivak2014,Bradley2018,FongSpivak2019}. Applied category theory (ACT) has become a field of its own, with dedicated conferences, a journal and software.

In this paper we propose that applied category theory is a well-suited language and toolbox for genomics, and we take first steps to support this claim. Our contributions are as follows.
\begin{itemize}
\item We review the development of genomics and identify three features of genomic data (heterogeneity, hierarchy and interconnection) that any integrative framework must handle (\cref{sec:genomics}).
\item We introduce the basic notions of category theory and the two central themes of ACT, compositionality and functorial semantics, and we match the features of genomic data with the categorical tools suited to them (\cref{sec:act}).
\item We present three bridges between ACT and genomics, through language, networks and ontology, summarising our earlier work on protein linguistics \citep{WuDisCoCat}, gene regulatory networks \citep{WuOpenPetri} and the Gene Ontology \citep{WuGeneOlogs}, each with a worked example, and we show that the three are instances of one pattern (\cref{sec:bridges}).
\item We construct from first principles a positional category of genes from a reference genome and show what it is good for: strand symmetry, synteny and chromosomal inversions, and nested chromatin domains (\cref{sec:genes}).
\item We sketch how these views can be integrated through a common database schema and list open problems (\cref{sec:integration}).
\end{itemize}

Our final aim, to uncover the organisational principles of genomes, is far beyond the reach of a single paper. What we hope to establish here is that the right tools exist and point in a clear direction. We call this programme \emph{categorical genomics}: the study of genomes with the methods of applied category theory. The paper is intended for both biologists and mathematicians; we therefore explain the biology and the mathematics we use as we go.

\section{Genomics: past, present and the next question}\label{sec:genomics}

Genomics studies all the genes of an organism together with their interactions with one another and with the environment, whereas genetics has traditionally focused on the function and inheritance of individual genes. The boundary between the two is not sharp, and both are closely intertwined with molecular biology and evolution. In the brief history below we therefore include developments that are not genomics in the narrow sense but that shaped it.

\subsection{A brief history}\label{sec:history}

The origins of genomics go back to Gregor Mendel, who presented his laws of inheritance in 1865 and published them in 1866 \citep{Mendel1866}. Before Mendel it was widely believed that the traits of the parents were blended in their offspring. Mendel showed instead that hereditary factors are passed on as discrete units, one from each parent. His work went largely unnoticed until it was rediscovered independently by three botanists in 1900 \citep{Posner1968}. The word ``gene'' for such a unit was coined by Wilhelm Johannsen in 1909 \citep{Johannsen1909}, long before its physical nature was known. In 1910, Thomas Hunt Morgan found a white-eyed male among his red-eyed fruit flies (\emph{Drosophila melanogaster}). Following its inheritance, he showed that the trait is linked to the X chromosome, the first assignment of a specific trait to a specific chromosome \citep{Morgan1910,Miko2008}.

For several decades afterwards most scientists believed that genetic information is carried by proteins rather than by DNA. The main reason was intuitive: DNA is built from only four kinds of nucleotides (A, C, G and T), whereas proteins are built from twenty kinds of amino acids, and proteins therefore seemed better suited to storing a vast amount of information. Experiments by \citet{Avery1944} and by \citet{Hershey1952} established that DNA is the genetic material. In retrospect, the lesson is that the number of basic components matters less than the ways in which they can be combined, a lesson that will recur in this paper.

\begin{figure}[t]
\centering
\includegraphics[width=0.42\textwidth]{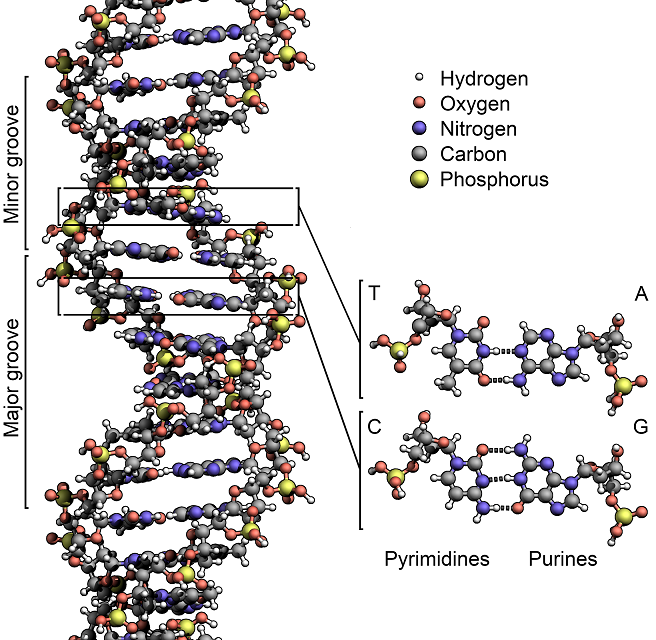}
\caption{The double-helix structure of DNA, with the complementary base pairs A--T and C--G (image by Zephyris, Wikimedia Commons).}
\label{fig:dna}
\end{figure}

How genes are stored in DNA and how DNA is copied remained unknown until 1953, when James Watson and Francis Crick, using X-ray diffraction data obtained by Rosalind Franklin and Raymond Gosling \citep{Franklin1953} and by Maurice Wilkins, proposed the double-helix structure of DNA \citep{Watson1953} (\cref{fig:dna}). Their one-page paper showed not only what DNA looks like but also, through complementary base pairing, how it could be replicated. It founded modern molecular biology.

How the information in DNA is used was worked out over the following decade. Crick formulated the ``central dogma'' that information flows from DNA to RNA to protein \citep{Crick1958}. \citet{Crick1961} showed that amino acids are encoded by triplets of bases, called codons; \citet{Nirenberg1961} identified the first codon; and by 1966 all 64 codons had been assigned to the twenty amino acids or to stop signals \citep{Nirenberg1963}.

The chain-termination method of DNA sequencing invented by \citet{Sanger1977} opened a new chapter. Sequencing technology advanced rapidly \citep{Behjati2013}, and together with the growth of bioinformatics, databases and genome browsers \citep{Hubbard2002,Kent2002}, it led to the genome sequence of \emph{Drosophila melanogaster}, the species of Morgan's white-eyed fly, in 2000 \citep{Adams2000}, and to the first drafts of the human genome sequence in 2001 \citep{Lander2001,Venter2001}.

\subsection{Genomics today}\label{sec:present}

The sequencing of whole genomes did not complete genomics; it transformed existing disciplines and opened new ones. Sequencing itself has continued to improve: the first truly complete, telomere-to-telomere sequence of a human genome was published in 2022 \citep{Nurk2022}, and a draft human \emph{pangenome}, which represents the variation among many individuals rather than a single reference, followed in 2023 \citep{Liao2023}. For \emph{Drosophila}, the Release~6 reference assembly is the basis of current annotation \citep{Hoskins2015}.

\paragraph{Medicine and genome editing.} Individual genome sequences allow some disease risks to be estimated before the disease develops, which is one of the aims of personalised medicine \citep{Khromykh2015,Suwinski2019}. Genome editing has advanced from zinc-finger nucleases and TALENs to CRISPR--Cas systems \citep{Gupta2014} and to base and prime editing, which change single bases without cutting both DNA strands \citep{Komor2016,Anzalone2019}. In 2023 the first CRISPR-based therapy, an edited-cell treatment for sickle-cell disease and $\beta$-thalassaemia whose first clinical results were reported by \citet{Frangoul2021}, was approved for clinical use.

\paragraph{Synthetic genomics.} Synthetic biology aims to redesign and build genomes. The international Sc2.0 project has constructed synthetic versions of all sixteen chromosomes of baker's yeast and is consolidating them into a single strain \citep{Zhao2023}.

\paragraph{Transcriptomics.} RNA sequencing (RNA-seq) measures how genes are expressed \citep{Wang2009}. Single-cell RNA-seq resolves expression in individual cells \citep{Macosko2015,Zheng2017} and has made possible cell atlases of whole organisms, including an atlas of the adult fruit fly \citep{Li2022FCA}.

\paragraph{Computation.} The volume of data has made genomics dependent on computation. Large public databases and genome browsers such as GenBank, RefSeq, Ensembl and the UCSC Genome Browser are maintained by institutes worldwide. Deep learning is now used throughout genomics \citep{Zou2019}; protein structure can be predicted with high accuracy \citep{Jumper2021}; and language models trained on protein and DNA sequences learn representations that capture structure and function \citep{Lin2023,DallaTorre2025,Nguyen2024}.

\paragraph{The 3D genome.} Although a DNA molecule is a linear sequence, chromosomes in the nucleus fold into complex and dynamic three-dimensional configurations that play important roles in gene regulation \citep{Bonev2016}. Chromosome conformation capture methods such as Hi-C \citep{LiebermanAiden2009,Kempfer2020} have shown that the genome is organised hierarchically: into large compartments and into smaller topologically associating domains (TADs), within which regulatory contacts are concentrated \citep{Dixon2012,Sexton2012}, and which can contain sub-domains nested inside them \citep{Gibcus2013,An2019}. How these domains form, for example by loop extrusion \citep{Fudenberg2016}, and how they affect genome function are being actively investigated \citep{Rowley2018}.

\subsection{The next question: integration}\label{sec:next}

As our picture of the 3D genome becomes clearer, the next question comes into view: what are the \emph{organisational principles} of a genome, structurally and functionally? Answering it would mean understanding genomes well enough to guide genome editing, genome design and the prediction of disease far more reliably than is possible today. To reach that level of understanding, the insights from the 3D genome must be combined with sequence and expression data and with the functional and relational knowledge accumulated in genomics. Three features of genomic data make this combination hard.

\paragraph{Heterogeneity.} Genomic data are diverse and heterogeneous. There are an estimated 8.7 million eukaryotic species, each with its own genome \citep{Mora2011}. For a single species, high-throughput technologies produce data of many types, such as sequence variants, chromatin accessibility, DNA methylation, histone modifications, chromatin contacts and RNA expression, each with its own formats, coordinate systems and noise \citep{Masseroli2019}.

\paragraph{Hierarchy.} Genomic data are hierarchical and multi-scale. Nucleotides form genes; genes are grouped in chromatin domains, which are nested within larger domains and compartments; and chromosomes form the genome. At the functional level, residues form protein domains, domains form proteins, proteins form complexes and pathways, and pathways form cellular processes.

\paragraph{Interconnection.} Genomic entities are densely interconnected. Distant chromatin segments contact one another in a coordinated but dynamic way \citep{Dixon2012}; transcription factors regulate their target genes, forming gene regulatory networks; and proteins interact in signalling pathways that cross-talk with one another. Moreover, the relationships themselves are related: one regulatory interaction may depend on another, and knowledge about one gene is transferred to another through orthology.

A framework for integrating genomic data should therefore (i) relate data of different kinds through maps that preserve their structure, (ii) build large systems from smaller parts, across scales, in a controlled way, and (iii) express relationships, and relationships between relationships, precisely. In the next section we explain why category theory provides exactly these capabilities.

\section{Applied category theory}\label{sec:act}

Category theory has a reputation for being highly abstract, even among mathematicians. Its basic notions are, however, simple, and its abstraction is what allows it to describe the common structure of very different systems. In this section we introduce the notions used in the rest of the paper, the two central themes of applied category theory, and its applications in biology, and we explain why it suits genomics.

\subsection{Categories, functors and natural transformations}\label{sec:basics}

\begin{definition}[Category; see \citealp{Awodey2010,Spivak2014}]\label{def:category}
A \emph{category} $\cat{C}$ consists of
\begin{itemize}
\item a collection of \emph{objects} $X, Y, Z, \dots$;
\item for each pair of objects $X, Y$, a collection of \emph{morphisms} $f\colon X \to Y$; $X$ is the \emph{domain} of $f$ and $Y$ its \emph{codomain};
\item for morphisms $f\colon X\to Y$ and $g\colon Y \to Z$, a \emph{composite} $g\circ f\colon X \to Z$;
\item for each object $X$, an \emph{identity} morphism $1_X\colon X \to X$,
\end{itemize}
such that composition is \emph{associative}, $h\circ(g\circ f) = (h\circ g)\circ f$, and \emph{unital}, $f\circ 1_X = f = 1_Y \circ f$, whenever these composites are defined.
\end{definition}

Objects represent things and morphisms represent ways of going from one thing to another: processes, relationships, transformations or ways of viewing. The essential feature is composition: two morphisms that meet at an object can be composed into one. Familiar examples include the category $\cat{Set}$ of sets and functions, and the category $\cat{FVect}$ of finite-dimensional real vector spaces and linear maps.

\begin{example}[Preorders as categories]\label{ex:preorder}
A \emph{preorder} on a set $S$ is a relation $\le$ that is reflexive ($s \le s$) and transitive (if $s\le s'$ and $s' \le s''$ then $s \le s''$). It is a \emph{partial order} if it is also antisymmetric (if $s \le s'$ and $s'\le s$ then $s=s'$), and a \emph{linear order} if in addition any two elements are comparable \citep[Def.~4.4.1.1]{Spivak2014}. Every preorder is a category: the objects are the elements of $S$, and there is exactly one morphism $s\to s'$ if $s\le s'$ and none otherwise. Reflexivity provides the identities and transitivity provides composition; associativity and unitality hold automatically, because there is at most one morphism between any two objects \citep[Prop.~5.2.1.13]{Spivak2014}. Such categories are called \emph{thin}. We use this example in \cref{sec:genes}.
\end{example}

A category can be related to another by a map that preserves its structure.

\begin{definition}[Functor]\label{def:functor}
A \emph{functor} $F\colon \cat{C}\to\cat{D}$ sends each object $X$ of $\cat{C}$ to an object $F(X)$ of $\cat{D}$ and each morphism $f\colon X\to Y$ to a morphism $F(f)\colon F(X)\to F(Y)$, such that $F(g\circ f) = F(g)\circ F(f)$ and $F(1_X) = 1_{F(X)}$.
\end{definition}

A functor between two preorders is simply an order-preserving (monotone) map \citep[Def.~4.4.4.1]{Spivak2014}. Functors can themselves be compared: a \emph{natural transformation} $\alpha\colon F\Rightarrow G$ between functors $F,G\colon \cat{C}\to\cat{D}$ consists of a morphism $\alpha_X\colon F(X)\to G(X)$ for each object $X$, such that $G(f)\circ\alpha_X = \alpha_Y\circ F(f)$ for every $f\colon X\to Y$. Categories, functors and natural transformations thus give a language for things, relationships between things, and relationships between relationships.

Two further constructions are used repeatedly below. Every category $\cat{C}$ has an \emph{opposite} category $\cat{C}^{\op}$ with the same objects and with all morphisms reversed \citep[Def.~6.2.1.1]{Spivak2014}; for a preorder, $\cat{C}^\op$ is the opposite order. And many categories carry a second, ``parallel'' form of composition: a \emph{monoidal category} has a tensor product $\otimes$ that combines objects $A\otimes B$ and morphisms $f\otimes g$ side by side, compatibly with sequential composition $\circ$ \citep{FongSpivak2019}. Monoidal categories have an intuitive graphical calculus of \emph{string diagrams}, in which objects are wires, morphisms are boxes, $\circ$ is vertical stacking and $\otimes$ is horizontal juxtaposition \citep{CoeckeKissinger2017}. Finally, \emph{limits} and \emph{colimits}, such as products, pullbacks, coproducts (disjoint unions) and pushouts (gluing), construct new objects from diagrams of old ones by universal properties \citep[Ch.~6]{Spivak2014}.

The power of these few notions comes from their generality. Through them, constructions from one field can be transported to another, and category theory has been proposed as an alternative foundation of mathematics \citep{Lawvere1966}. They also reveal deep analogies between apparently unrelated subjects; for example, physics, topology, logic and computation share the structure of closed symmetric monoidal categories \citep{BaezStay2011}.

\subsection{Two themes of applied category theory}\label{sec:themes}

Two themes recur throughout applied category theory \citep{Bradley2018}.

\paragraph{Compositionality.} A system is compositional if its behaviour is determined by the behaviour of its parts and by the way in which the parts are combined. This is Frege's principle for the meaning of sentences, and it is also the name of the journal of the field, \emph{Compositionality} \citep{Compositionality}. Categories make the idea precise: ``combining'' is composition $\circ$ (in series) or tensor $\otimes$ (in parallel), and ``open'' systems that can be connected to others are modelled as morphisms whose domain and codomain are their interfaces. A standard way of building such morphisms is the \emph{cospan} $X \to S \leftarrow Y$, an object $S$ with two boundary maps; two cospans are composed by gluing along the shared boundary, which is a pushout. Decorated and structured cospans \citep{Fong2015,BaezCourser2020} provide general machinery of this kind, and they underlie the open Petri nets of \cref{sec:network}.

\paragraph{Functorial semantics.} The idea, going back to \citet{Lawvere1963}, is to separate the \emph{syntax} of a system (how its descriptions are put together) from its \emph{semantics} (what the descriptions mean), and to model meaning as a functor
\[
F\colon \text{Syntax} \longrightarrow \text{Semantics}.
\]
Because $F$ preserves composition, $F(g\circ f) = F(g)\circ F(f)$ (and, for monoidal functors, $F(f\otimes g) \cong F(f)\otimes F(g)$), the meaning of a composite is obtained by composing the meanings of the parts. Functorial semantics is therefore how compositionality is achieved in practice. Grammar and meaning in natural language are the motivating example \citep{Bradley2018}, and we will see that the same pattern underlies all three bridges of \cref{sec:bridges}.

\paragraph{The field.} Spivak's book \emph{Category Theory for the Sciences} \citep{Spivak2014} introduced ologs and categorical databases as tools for scientists. The first Applied Category Theory school and workshop in 2018 \citep{ACT2018} and the journal \emph{Compositionality} \citep{Compositionality} marked the emergence of ACT as a discipline. Applications now include quantum computing, natural language processing, programming languages, network theory, databases, chemistry, epidemiology, dynamical systems, game theory, probability and cybernetics \citep{Bradley2018,FongSpivak2019}. ACT also has a growing body of software, such as the AlgebraicJulia ecosystem for compositional scientific modelling \citep{Patterson2022,Libkind2022} and the \texttt{lambeq} library for compositional language processing \citep{Kartsaklis2021}.

\subsection{Category theory in biology}\label{sec:ct-biology}

The theoretical biologist Robert Rosen was among the first to use category theory in biology \citep{Rosen1958a,Rosen1958b}. Asking ``what is life?'', he argued that reductionism is inadequate for the study of the functional organisation of organisms, and he modelled an organism as a system whose organisation, rather than its material parts, is the object of study. His metabolism--repair, or $(M,R)$, systems and the resulting ``relational biology'' used category-theoretic methods \citep{Rosen1991,Baianu2006}.

Other work has focused on the molecular level. \citet{CarboneGromov2001} described the spatial structure of DNA, RNA and proteins with topological surfaces and spaces, a related but not categorical approach. \citet{Sawamura2014} described algebraic features of DNA base sequences with a wallpaper symmetry group, aiming towards a category-theoretic systematisation of molecular biology. \citet{Tuyeras2017,Tuyeras2018a,Tuyeras2018b} developed a categorical framework for genetics in which sequence alignment, mutation, homologous recombination, haplotypes, CRISPR editing and genetic linkage are formalised. \citet{Spivak2011} used ologs to compare the hierarchical structure of protein materials with that of social networks, and \citet{BaezPollard2017} gave a compositional framework for chemical reaction networks, which applies directly to biochemical systems.

In neuroscience, \citet{Ehresmann2007} used colimits to model how complex hierarchical systems such as the brain are organised and evolve, building on proposals to use higher-dimensional algebra for brain structure \citep{BrownPorter2003}. Process theories, which are symmetric monoidal categories with string-diagram reasoning \citep{CoeckeKissinger2017}, have even been used to model consciousness \citep{Signorelli2020,TullKleiner2020}.

Genomics, by contrast, has been touched only sporadically. Tuyéras's work addresses genetics and parts of sequence analysis, and our own earlier papers addressed protein linguistics, gene regulatory networks and the Gene Ontology \citep{WuDisCoCat,WuOpenPetri,WuGeneOlogs}. What has been missing is an explicit programme connecting ACT with genomics as a whole, which this paper aims to begin.

\subsection{Why applied category theory suits genomics}\label{sec:why}

The three features of genomic data identified in \cref{sec:next} correspond to three strengths of category theory (\cref{tab:features}).

\emph{Heterogeneity} calls for a way to relate different kinds of data without forcing them into one format. Functors do exactly this: they relate structures of different kinds while preserving what they have in common. In particular, a database schema can be read as a category and a database instance as a functor to $\cat{Set}$; a functor between schemas then induces canonical \emph{data migration} functors that move data between them \citep{Spivak2012}, \citep[Sec.~7.1.4]{Spivak2014}. This turns data integration from an ad hoc procedure into a mathematical construction.

\emph{Hierarchy} calls for a way to build systems from parts at several scales. Sequential and parallel composition in monoidal categories, gluing of open systems along interfaces, and operads, in which systems are nested inside larger systems as in wiring diagrams \citep[Sec.~7.4]{Spivak2014}, \citep{Spivak2013wd}, provide exactly such multi-scale composition, with a guarantee that the behaviour of the whole is computed from the behaviour of the parts.

\emph{Interconnection} calls for a precise language of relationships. Morphisms, spans (many-to-many relations), cospans (open systems) and universal constructions such as pullbacks and pushouts express relationships, and natural transformations express relationships between relationships.

\begin{table}[t]
\centering
\small
\begin{tabularx}{\textwidth}{>{\raggedright\arraybackslash}p{2.6cm} >{\raggedright\arraybackslash}X >{\raggedright\arraybackslash}p{4.2cm}}
\toprule
Feature of genomic data & Categorical tools & Where used in this paper\\
\midrule
Heterogeneity & functors; schemas as categories, instances as functors to $\cat{Set}$; data migration & Gene ologs (\cref{sec:ontology}); common schema (\cref{sec:integration})\\
Hierarchy, multiple scales & monoidal categories and string diagrams; cospans glued by pushout; operads & DisCoCat (\cref{sec:language}); open Petri nets (\cref{sec:network}); nested domains (\cref{sec:using})\\
Interconnection & morphisms, spans, natural transformations; limits and colimits & annotations as spans (\cref{sec:ontology}); positional category (\cref{sec:genes})\\
\bottomrule
\end{tabularx}
\caption{Features of genomic data and the categorical tools that address them.}
\label{tab:features}
\end{table}

\section{Three bridges between applied category theory and genomics}\label{sec:bridges}

\begin{figure}[t]
\centering
\includegraphics[width=0.95\textwidth]{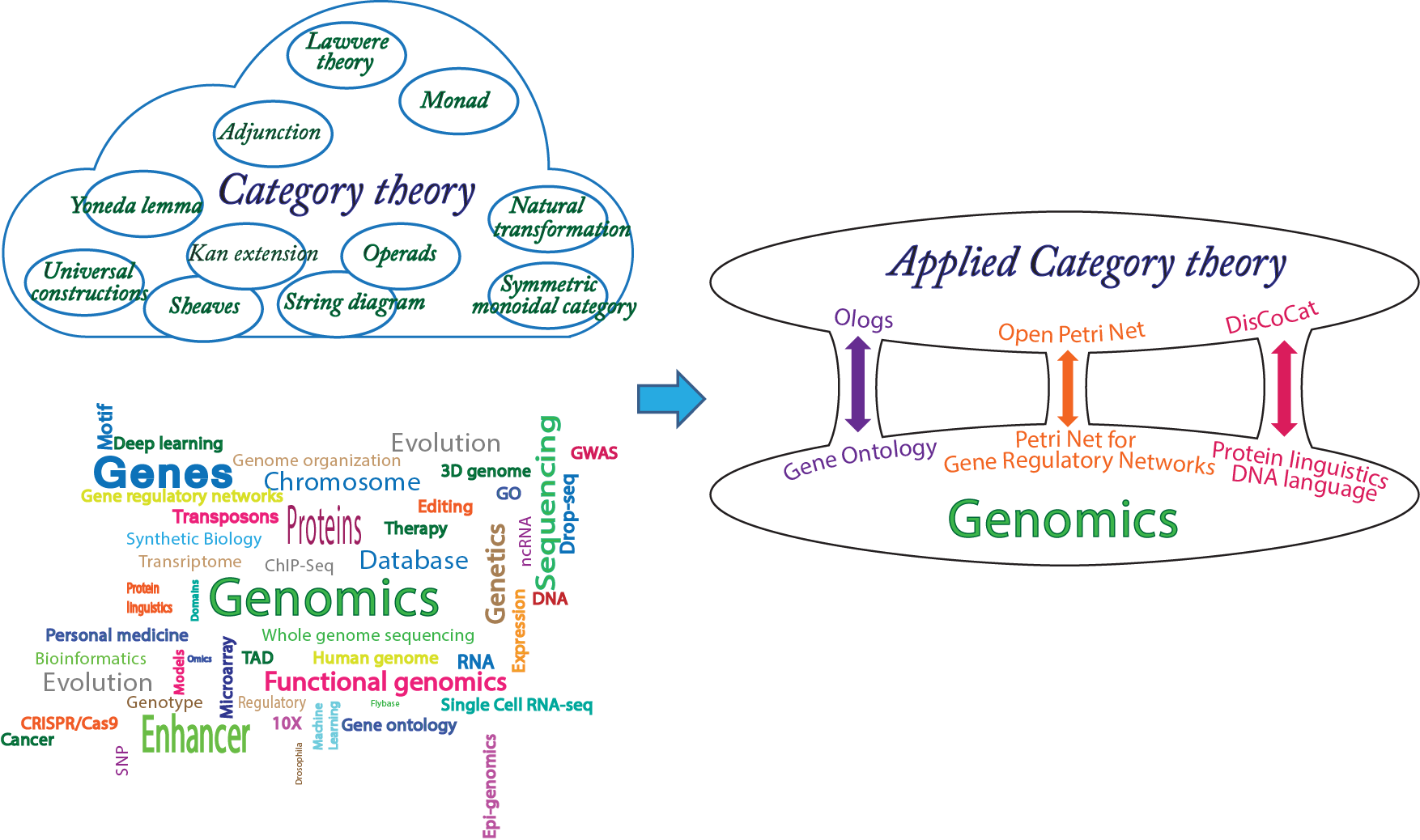}
\caption{Three bridges between applied category theory and genomics. Left: some concepts of category theory (top) and of genomics (bottom). Right: ologs connect with the Gene Ontology, open Petri nets with gene regulatory networks, and the DisCoCat model with protein and DNA linguistics.}
\label{fig:bridges}
\end{figure}

There are two ways to connect genomics with category theory: to use categorical models that already exist for related problems, and to build categorical structures for genomics from scratch. In this section we follow the first route; \cref{sec:genes} follows the second. \Cref{fig:bridges} shows three topics that serve as bridges between the fields: language, networks and ontology. For each we summarise the categorical model, its application to genomics as developed in our companion papers \citep{WuDisCoCat,WuOpenPetri,WuGeneOlogs}, a worked example and the main limitations. In \cref{sec:pattern} we show that the three bridges are instances of a single pattern.

\subsection{Language: a compositional model of protein function}\label{sec:language}

\paragraph{The language of life.} Since the genome is physically a long string over the alphabet $\{\mathrm{A},\mathrm{C},\mathrm{G},\mathrm{T}\}$, it is natural to view it as a text written in a language, and to borrow tools from linguistics. Following Chomsky's generative grammar, Searls and others developed formal grammars for genes, RNA secondary structure and protein structure \citep{Brendel1984,Searls1992,Searls2002,Chiang2006,Searls2013}. These grammars describe the \emph{structure} of macromolecules. The categorical model below aims also at their \emph{meaning}, that is, their function.

\paragraph{The DisCoCat model.} The categorical compositional distributional (DisCoCat) model of \citet{Coecke2010} computes the meaning of a sentence from the meanings of its words and its grammatical structure. Grammar is given by a \emph{pregroup} \citep{Lambek1999}: each word receives a type built from basic types (such as $n$ for noun and $s$ for sentence) and their left and right adjoints $x^l, x^r$, subject to the contractions $x^l x \to 1$ and $x x^r \to 1$. A transitive verb, for example, has type $n^r s\, n^l$, and the sentence ``Alice loves Bob'' is grammatical because its types reduce to the sentence type:
\[
n\,(n^r s\, n^l)\,n \;\longrightarrow\; s .
\]
Word meanings are vectors obtained from data, for example from the contexts in which words occur. The key observation of \citet{Coecke2010} is that pregroup grammars and vector spaces share a structure: both form compact closed (more precisely, rigid monoidal) categories, in which the contractions correspond to the ``cups'' $\epsilon$ that, for vector spaces, are inner products. Meaning is then given by a strong monoidal functor
\[
F\colon \cat{Grammar} \longrightarrow \cat{FVect}
\]
from the free compact closed category generated by the basic types, whose morphisms are grammatical reductions, to the category of finite-dimensional vector spaces \citep{Coecke2013}. $F$ sends each basic type to a vector space, so that a word of type $p$ has a meaning vector in $F(p)$, and each reduction $\alpha$ to a linear map $F(\alpha)$. The meaning of a sentence $w_1\cdots w_n$ is $F(\alpha)(\vec w_1\otimes\cdots\otimes\vec w_n)$. All sentences have meanings in the same space $S = F(s)$, so they can be compared by the cosine similarity $\langle \vec s\,|\,\vec s\,'\rangle / \lVert\vec s\rVert\lVert\vec s\,'\rVert$, whatever their grammatical structure. This is functorial semantics in its most literal form: grammar is the syntax and $F$ is the semantics \citep{Bradley2018}.

\paragraph{Proteins as sentences.} Proteins are the products of protein-coding genes and carry out most of their functions. They are built in a modular way from \emph{domains}: independently folding structural and functional units that are conserved in evolution, whereas the ways in which domains are combined in proteins vary greatly \citep{Gimona2006,Bashton2007}. This suggests an analogy in which a protein is a sentence, its domains are words, and the biological function of a domain or protein is its meaning \citep{WuDisCoCat}. The analogy is useful because the function of many domains is known, whereas predicting the function of a protein with a new combination of domains remains difficult \citep{Radivojac2013,Zhou2019}.

\paragraph{A pregroup grammar for proteins.} Domains are typed by broad functional class, for example $b$ (binding; with subtypes such as $b_2$ for DNA binding), $c$ (catalytic), $i$ (protein--protein interaction), $r$ (regulatory) and $t$ (channel or transporter), and there are protein types $p$ (protein), $p_1$ (single-domain protein) and $p_2$ (multi-domain protein), with order postulates such as $b_2\to b$ and $p_2\to p$. The linker regions between domains play the role of relational words such as transitive verbs: a linker between a domain of type $x$ and a domain of type $y$ has type
\[
L_{xy}\colon\; x^r p_2\, y^l, \qquad\text{so that}\qquad x\,(x^r p_2\, y^l)\,y \;\longrightarrow\; p_2 \;\longrightarrow\; p.
\]
The lexicon contains an entry $L_{xy}$ only for pairs of domain types observed in real proteins, so that grammaticality is a non-trivial check; only a small fraction of all possible domain combinations occur in nature, and these combinations have been found to follow a ``quasi-universal grammar'' \citep{Yu2019}. Proteins with $m\ge 3$ domains are handled by \emph{continuation linkers} $K_z\colon p_2^r p_2\, z^l$, which extend the protein assembled so far by one domain; by induction on $m$, the type of any protein $D_1 L_1 D_2 \cdots L_{m-1} D_m$ then reduces to $p$ \citep{WuDisCoCat}.

\paragraph{Semantics.} The functor $F$ sends every domain type to a domain space $D$ and every protein type to a protein space $P$. Domain vectors $u\in D$ can be obtained from sequence embeddings, such as ProtVec \citep{Asgari2015} or modern protein language models \citep{Lin2023}, applied to the domain's subsequence. A linker $L_{xy}$ is represented by a tensor $L\in D\otimes P\otimes D$, which is learned from data. Applying $F$ to the reduction of a two-domain protein with domain vectors $u$ and $w$ gives
\begin{equation}\label{eq:protein-meaning}
\vec v \;=\; (\epsilon_D\otimes 1_P\otimes \epsilon_D)(u\otimes L\otimes w) \;=\; \sum_j \bigl(u^{\top} L^{(j)} w\bigr)\, f_j \;\in P,
\end{equation}
where $\{f_j\}$ is a basis of $P$ and $L^{(j)}$ is the matrix of entries $L_{ijk}$ for fixed $j$. Each coordinate of the protein's meaning is thus a bilinear form in its two domain vectors (\cref{fig:discocat}).

\begin{figure}[t]
\centering
\begin{tikzpicture}[x=1cm,y=1cm, every node/.style={font=\small}]
  \draw[fill=blue!12] (-0.55,0) -- (0.55,0) -- (0,0.8) -- cycle; \node at (0,0.3) {$u$};
  \draw[fill=orange!18] (1.3,0) -- (3.7,0) -- (2.5,0.8) -- cycle; \node at (2.5,0.3) {$L_{xy}$};
  \draw[fill=blue!12] (4.45,0) -- (5.55,0) -- (5,0.8) -- cycle; \node at (5,0.3) {$w$};
  \draw[thick] (0,0) .. controls (0,-1.1) and (1.6,-1.1) .. (1.6,0);
  \draw[thick] (3.4,0) .. controls (3.4,-1.1) and (5,-1.1) .. (5,0);
  \draw[thick] (2.5,0) -- (2.5,-1.6) node[below] {$P$};
  \node[left] at (0,-0.25) {$D$}; \node[right] at (1.6,-0.25) {$D$};
  \node[left] at (3.4,-0.25) {$D$}; \node[right] at (5,-0.25) {$D$};
  \node at (0.8,-1.05) {$\epsilon_D$}; \node at (4.2,-1.05) {$\epsilon_D$};
  \node at (0,1.2) {domain $D_1$}; \node at (2.5,1.2) {linker $L_1$}; \node at (5,1.2) {domain $D_2$};
  \node at (0,1.65) {$x$}; \node at (2.5,1.65) {$x^r\, p_2\, y^l$}; \node at (5,1.65) {$y$};
  \node[align=left, anchor=west] at (6.6,0.3) {grammar (syntax):\\[2pt] $x\,(x^r p_2\, y^l)\,y \to p_2 \to p$\\[8pt] meaning (semantics):\\[2pt] $\vec v = \sum_j (u^{\top}L^{(j)}w)\,f_j\in P$};
\end{tikzpicture}
\caption{The DisCoCat model for a two-domain protein. Top row: the pregroup types of the two domains and the linker. The triangles are the meaning vectors of the ``words'': domain vectors $u, w\in D$ and the linker tensor $L_{xy}\in D\otimes P\otimes D$. The cups $\epsilon_D$ (inner products) are the images under $F$ of the pregroup contractions, and the one open wire carries the meaning of the protein in $P$. Adapted from \citet{WuDisCoCat}.}
\label{fig:discocat}
\end{figure}

\begin{example}[A toy protein]\label{ex:mockP}
Take $D=\R^3$ and $P=\R^2$, a regulatory domain $u=(30,12,9)^\top$ followed by a catalytic domain $w = (2,4,10)^\top$, and a linker tensor $L_{rc}$ with slices
\[
L^{(1)} = \begin{pmatrix}1&0&0\\0&1&0\\0&0&1\end{pmatrix}, \qquad L^{(2)} = \begin{pmatrix}0&1&0\\0&0&1\\1&0&0\end{pmatrix}.
\]
Then \eqref{eq:protein-meaning} gives $\vec v = (u^\top L^{(1)} w,\; u^\top L^{(2)} w)^\top = (198,\,258)^\top$. Swapping the two domains gives $(198,\,360)^\top$, because $L^{(2)}$ is not symmetric, whereas the ``bag-of-words'' sum $u+w = (32,16,19)^\top$ cannot distinguish the two orders. The model therefore respects domain order, as grammar respects word order. (With $P=\R$ and $L$ the identity, the ``meaning'' would reduce to the inner product $\langle u|w\rangle = 198$, which measures the similarity of the two domains rather than the function of the protein; the order-three linker tensor is what gives proteins a meaning space of their own.)
\end{example}

For real proteins, the \emph{Drosophila} transcription factor FoxP, one of whose human orthologues, FOXP2, is mutated in a severe speech and language disorder \citep{Lai2001}, has a coiled-coil region (typed $r$) followed by a forkhead DNA-binding domain ($b_2$). The fly sex-determination protein Fruitless has the same architecture $(r, b_2)$, with a BTB domain and zinc-finger domains, and therefore uses the same linker entry $L_{rb_2}$; the two proteins differ in the model only through their domain vectors, and their functional similarity is the cosine similarity of their composed vectors \citep{WuDisCoCat}.

\paragraph{Assumptions, testing and outlook.} The model rests on two hypotheses that must be tested rather than assumed: domains that occur in similar contexts have similar functions (a distributional hypothesis), and the function of a protein is, to a useful approximation, a multilinear function of the functions of its domains, determined by its architecture (a compositional hypothesis). Because \eqref{eq:protein-meaning} is linear in $L$, linker tensors can be estimated by regression from proteins with known functions, for example using Gene Ontology annotations as targets, and the model can be compared with additive baselines and with whole-protein embeddings on held-out proteins \citep{WuDisCoCat}. DisCoCat has been extended from sentences to texts by DisCoCirc \citep{Coecke2021}, in which the meanings of entities evolve as sentences act on them. A protein analogue would describe how the proteins of a complex or pathway modify one another's functions. This is the natural next level of composition, from proteins to the interacting gene products encoded by a genome; it does not, however, follow automatically from the protein model, and it requires its own grammar of interactions.

\subsection{Networks: gene regulatory networks as open Petri nets}\label{sec:network}

\paragraph{Gene regulatory networks and Petri nets.} A gene regulatory network (GRN) describes how genes regulate one another's expression: gene $X$ regulates gene $Y$ if a change in the expression of $X$ induces a change in the expression of $Y$, by activation or by inhibition. GRNs have been modelled with Boolean networks, Bayesian networks, differential equations and Petri nets, among other formalisms \citep{deJong2002,Delgado2019}. Petri nets \citep{Reisig1985,Murata1989} combine a precise mathematical theory with an intuitive graphical notation and have been used for GRNs for some time \citep{Chaouiya2004,Chaouiya2007,Steggles2007}.

\begin{definition}[Petri net; \citealp{Meseguer1990,BaezMaster2020}]
A \emph{Petri net} consists of a set $S$ of \emph{places}, a set $T$ of \emph{transitions}, and two functions $s,t\colon T\to\N[S]$, where $\N[S]$ is the set of finite formal sums of places with natural-number coefficients. A \emph{marking} $m\in\N[S]$ assigns a number of tokens to each place. A transition $\tau$ is enabled at $m$ if $m\ge s(\tau)$, and \emph{firing} it produces the marking $m - s(\tau) + t(\tau)$.
\end{definition}

For a GRN, places are genes (more precisely, their products) and tokens are units of gene product. Each regulatory interaction becomes a transition $T_{g_1,g_2}$ (target $g_1$, regulator $g_2$), with a choice of whether the regulator is consumed \citep{Chaouiya2004,WuOpenPetri}:
\begin{center}
\small
\begin{tabular}{lll}
\toprule
interaction & $s(T_{g_1,g_2})$ & $t(T_{g_1,g_2})$\\
\midrule
$g_2$ activates $g_1$, $g_2$ not consumed & $g_2$ & $g_1+g_2$\\
$g_2$ activates $g_1$, $g_2$ consumed & $g_2$ & $g_1$\\
$g_2$ inhibits $g_1$, $g_2$ not consumed & $g_1+g_2$ & $g_2$\\
degradation $d_g$ of $g$ & $g$ & $0$\\
\bottomrule
\end{tabular}
\end{center}
Here inhibition is modelled as regulator-dependent removal of the target's product, and degradation transitions prevent the unbounded accumulation of products that non-consuming activation would otherwise cause.

\paragraph{The problem of composition.} An ordinary Petri net describes one closed system. \citet{Meseguer1990} showed that the behaviours of a net form a monoidal category whose objects are markings and whose morphisms are processes, built from transitions by sequential composition $\circ$ and parallel composition $\oplus$. This composes processes \emph{within} one net, but it does not compose nets: the only way to combine two nets in that framework is to place them side by side, without interaction. A GRN, however, is assembled from sub-networks that \emph{share} genes: the target in one module is a regulator in another. What is needed is a way to glue nets along common places.

\paragraph{Open Petri nets.} \citet{BaezMaster2020} make this precise. An \emph{open Petri net} $P\colon X\nrightarrow Y$ is a Petri net with places $S$ together with functions $i\colon X\to S$ and $o\colon Y\to S$ that mark certain places as inputs and outputs, that is, a cospan of Petri nets $LX\to P\leftarrow LY$, where $LX$ is the net with places $X$ and no transitions. Given $P\colon X\nrightarrow Y$ and $Q\colon Y\nrightarrow Z$, the composite $Q\odot P\colon X\nrightarrow Z$ glues the outputs of $P$ to the inputs of $Q$; formally it is a pushout in the category of Petri nets. Concretely, places that are images of the same point of $Y$ are identified, transitions are the disjoint union of those of $P$ and $Q$, and arc weights add when places are merged. Open Petri nets, with sets as objects, form a symmetric monoidal (double) category $\mathsf{Open}(\cat{Petri})$, in which $\odot$ is sequential and disjoint union is parallel composition \citep{BaezMaster2020,BaezCourser2020}.

\paragraph{Semantics.} An open Petri net is syntax, and its behaviour is given by functors out of $\mathsf{Open}(\cat{Petri})$ \citep{BaezMaster2020,BaezPollard2017,WuOpenPetri}. (i) The \emph{operational} semantics extends the Meseguer--Montanari construction to open nets, so that the processes of a composite net are computed from those of its parts. (ii) The \emph{reachability} semantics sends a net to the relation between input and output markings that it can realise; it is only \emph{lax} functorial, $\blacksquare Q\circ\blacksquare P\subseteq\blacksquare(Q\odot P)$, because in a composite net tokens may pass back and forth between the parts. (iii) The \emph{rate-equation} semantics of \citet{BaezPollard2017} attaches a rate constant to each transition and sends a net to its mass-action differential equation with inflows and outflows at the boundary; this assignment is functorial, so the dynamics of a composite are obtained by gluing the dynamics of its parts.

\paragraph{Rules for modules.} The composition law implies three simple rules for cutting a GRN into modules so that composition gives back the original network \citep{WuOpenPetri}: (R1) cut through genes, not through interactions, so that every gene taking part in interactions in two modules lies in their shared interface; (R2) assign each transition, including the degradation of a shared gene, to exactly one module, since transitions are never identified; and (R3) treat the input/output direction as bookkeeping only, so that feedback between modules needs no special treatment.

\begin{figure}[t]
\centering
\begin{tikzpicture}[x=1cm,y=1cm]
  \node at (-0.4,2.0) {(a)};
  \node[font=\small] (gA) at (0,0) {$A$};
  \node[font=\small] (gB) at (1.3,0.9) {$B$};
  \node[font=\small] (gC) at (1.3,-0.9) {$C$};
  \node[font=\small] (gD) at (2.6,0.9) {$D$};
  \node[font=\small] (gE) at (2.6,-0.9) {$E$};
  \draw[arr] (gA) -- (gB); \draw[arr] (gA) -- (gC);
  \draw[arr] (gB) -- (gD); \draw[arr] (gC) -- (gE);
  \draw[thick,-{Bar[width=3mm]}] (gC) -- (gD);
  \begin{scope}[xshift=4.4cm]
  \node at (-1.1,2.35) {(b)};
  \node[place] (A) at (0,0) {$A$};
  \node[trans] (TBA) at (1.5,1.0) {};
  \node[trans] (TCA) at (1.5,-1.0) {};
  \node[place] (B) at (3.2,1.0) {$B$};
  \node[place] (C) at (3.2,-1.0) {$C$};
  \node[trans] (TDB) at (4.9,1.0) {};
  \node[trans] (TDC) at (4.9,0) {};
  \node[trans] (TEC) at (4.9,-1.0) {};
  \node[place] (D) at (6.6,1.0) {$D$};
  \node[place] (E) at (6.6,-1.0) {$E$};
  \node[font=\scriptsize] at (1.5,1.6) {$T_{B,A}$};
  \node[font=\scriptsize] at (1.5,-1.6) {$T_{C,A}$};
  \node[font=\scriptsize] at (4.9,1.6) {$T_{D,B}$};
  \node[font=\scriptsize] at (5.45,-0.3) {$T_{D,C}$};
  \node[font=\scriptsize] at (4.9,-1.6) {$T_{E,C}$};
  \draw[{Stealth[length=2mm]}-{Stealth[length=2mm]},thick] (A) -- (TBA);
  \draw[{Stealth[length=2mm]}-{Stealth[length=2mm]},thick] (A) -- (TCA);
  \draw[arr] (TBA) -- (B); \draw[arr] (TCA) -- (C);
  \draw[{Stealth[length=2mm]}-{Stealth[length=2mm]},thick] (B) -- (TDB);
  \draw[arr] (TDB) -- (D);
  \draw[{Stealth[length=2mm]}-{Stealth[length=2mm]},thick] (C) -- (TDC);
  \draw[arr] (D) -- (TDC);
  \draw[arr] (C) -- (TEC); \draw[arr] (TEC) -- (E);
  \begin{scope}[on background layer]
    \fill[blue!6, rounded corners] (-0.6,-1.95) rectangle (3.75,1.95);
    \draw[blue!60, dashed, thick, rounded corners] (-0.6,-1.95) rectangle (3.75,1.95);
    \fill[red!6, fill opacity=0.7, rounded corners] (2.65,-2.05) rectangle (7.15,2.05);
    \draw[red!60, dashed, thick, rounded corners] (2.65,-2.05) rectangle (7.15,2.05);
  \end{scope}
  \node[font=\small, blue!60!black] at (1.0,2.35) {module $P\colon X\nrightarrow Y$};
  \node[font=\small, red!60!black] at (5.4,2.35) {module $Q\colon Y\nrightarrow Z$};
  \node[font=\scriptsize] at (3.2,-2.35) {interface $Y=\{B,C\}$};
  \end{scope}
\end{tikzpicture}
\caption{(a) A small gene regulatory network: $A$ activates $B$ and $C$, $B$ activates $D$, $C$ inhibits $D$ and activates $E$. (b) Its Petri net as the composite $Q\odot P$ of two open Petri net modules glued along the shared genes $B$ and $C$. Circles are places (genes), grey rectangles are transitions (interactions); double-headed arrows are read arcs (the regulator is not consumed); only $C$ is consumed, when it activates $E$. Degradation transitions $d_A,\dots,d_E$ are not drawn; $d_B$ and $d_C$ belong to $P$. Adapted from \citet{WuOpenPetri}.}
\label{fig:petri}
\end{figure}

\begin{example}[Composing a GRN from two modules]\label{ex:grn}
Consider the GRN of \cref{fig:petri}(a). The upstream module $P\colon X\nrightarrow Y$ has places $\{A,B,C\}$, transitions $T_{B,A}\colon A\to A+B$, $T_{C,A}\colon A\to A+C$ and degradations $d_A, d_B, d_C$, an input at $A$, and outputs $Y=\{B,C\}$. The downstream module $Q\colon Y\nrightarrow Z$ has places $\{B',C',D,E\}$, transitions $T_{D,B}\colon B'\to B'+D$, $T_{D,C}\colon C'+D\to C'$, $T_{E,C}\colon C'\to E$ and degradations $d_D, d_E$, inputs $Y$ at $B', C'$, and outputs at $D$ and $E$. The composite $Q\odot P$ identifies $B$ with $B'$ and $C$ with $C'$; its transitions are the ten transitions of $P$ and $Q$, which are exactly those of the Petri net of the whole GRN with one degradation per gene (\cref{fig:petri}(b)). With rate constants $k$ for the regulatory transitions and $\delta$ for the degradations, the open rate equations of $P$ and $Q$ glue, after the flows across $Y$ cancel, into the mass-action dynamics of the whole network:
\[
\begin{aligned}
\dot A &= -\delta_A A + I_x, &
\dot B &= k_{BA}A - \delta_B B, &
\dot C &= k_{CA}A - (k_{EC}+\delta_C)\,C,\\
\dot D &= k_{DB}B - k_{DC}\,C D - \delta_D D - O_{z_1}, &
\dot E &= k_{EC}C - \delta_E E - O_{z_2}, &&
\end{aligned}
\]
where $I_x$ is an inflow at the input $A$ and $O_{z_1}, O_{z_2}$ are outflows at the outputs $D$ and $E$ \citep{WuOpenPetri}. Thus both the network and its dynamics are computed from those of its modules.
\end{example}

\paragraph{Limitations.} The token-based encoding is coarse: it does not distinguish transcription from translation, and inhibitor arcs, used in some Petri net models of regulation, lie outside the framework. Reachability of a composite can be bounded from below, but not in general computed, from the reachability of its parts. And mass-action kinetics is only one of many possible rate laws. The value of the approach is that it makes the modular structure of GRNs explicit and mathematically sound, and that the same syntax supports several semantics.

\subsection{Ontology: Gene ologs}\label{sec:ontology}

\paragraph{The Gene Ontology.} To transfer knowledge of gene function between species, the Gene Ontology (GO) provides a controlled vocabulary for annotating genes and gene products \citep{Ashburner2000}. GO consists of three disjoint namespaces: biological process, molecular function and cellular component. It has become a central resource for the unification of biology: as of the release of 19 June 2026, it contained 38,263 terms and 14,736,040 annotations for 1,750,636 gene products \citep{GO2026,WuGeneOlogs}. GO is represented in the Web Ontology Language (OWL), a description logic that supports automated reasoning \citep{Hastings2017,Hitzler2010}. To keep reasoning tractable at the scale of GO, restricted profiles such as OWL~2~EL are used, which omit constructs such as class union \citep{Motik2012}. GO-CAM models extend standard annotations by linking activities, the gene products that enable them, the processes they are causally upstream of, and the locations in which they occur \citep{Thomas2019}.

\paragraph{Ologs.} An \emph{olog} (ontology log) is a category presented in natural language, introduced by \citet{SpivakKent2012} as a categorical framework for knowledge representation \citep[Sec.~2.3]{Spivak2014}. Its objects, called \emph{types}, are drawn as boxes containing indefinite noun phrases, such as \type{a gene}; each type stands for a set of things. Its morphisms, called \emph{aspects}, are labelled arrows such as \type{a gene} $\xrightarrow{\text{is}}$ \type{a segment of DNA}. Aspects must be \emph{functional}: each instance of the source is related to exactly one instance of the target. This is what allows them to be composed. A \emph{fact} declares that two paths with the same source and target are equal, which is what turns a graph into a category; it is marked with a check mark. Instance data on an olog $\cat{C}$ is a functor $I\colon\cat{C}\to\cat{Set}$, which assigns to each type a set (a table of instances) and to each aspect a function (a column), such that all facts hold; an olog is therefore also a database schema \citep{SpivakKent2012,Spivak2012}. Finally, types can be constructed from others by limits and colimits, such as products, pullbacks (``and''), coproducts and pushouts (``or'' without double counting), whose meaning is fixed by universal properties.

\paragraph{Gene ologs.} In a Gene olog \citep{WuGeneOlogs}, most types come from GO terms, read as the set of \emph{instances} of the term (particular occurrences of a process or activity, particular cellular structures), while gene products form a separate type. Writing a Gene olog forces a decision, for each GO relation, about whether it is functional and can be an aspect, or many-to-many and must be represented as a \emph{span}: a new type whose instances are the related pairs, with one aspect to each participant. \Cref{fig:olog}(a) shows a small olog relating genes, transcripts and protein isoforms. It also illustrates a modelling rule: the arrows point \emph{from} transcripts and isoforms \emph{to} genes, because a gene may have several transcripts and isoforms (or none), so arrows starting at \type{a gene}, such as ``a gene codes for a protein'', would not be functional. A particular gene such as FoxP is not a type but an instance, an element of $I(\type{a gene})$.

\begin{figure}[t]
\centering
\begin{tikzpicture}[x=1cm,y=1cm]
  \node at (-1.2,1.2) {(a)};
  \node[olog] (tr) at (0,0) {a protein-coding\\ transcript};
  \node[olog] (iso) at (6.6,0) {a protein isoform};
  \node[olog] (gene) at (3.3,-2.2) {a gene};
  \node[olog] (dna) at (7.6,-2.2) {a segment of DNA};
  \draw[arr] (tr) -- node[above,font=\scriptsize] {has as translation product} (iso);
  \draw[arr] (tr) -- node[left,font=\scriptsize,pos=0.45] {is transcribed from\ } (gene);
  \draw[arr] (iso) -- node[right,font=\scriptsize,pos=0.45] {\ is encoded by} (gene);
  \draw[arr,{Hooks[right]}-{Stealth[length=2.2mm]}] (gene) -- node[below,font=\scriptsize] {is} (dna);
  \node at (3.3,-0.8) {\large$\checkmark$};
\end{tikzpicture}
\\[10pt]
\begin{tikzpicture}[x=1cm,y=1cm]
  \node at (-2.6,1.0) {(b)};
  \node[olog] (ann) at (0,0) {an annotation $(g,r,t)$};
  \node[olog] (gp) at (-3.6,-1.6) {a gene product};
  \node[olog] (rel) at (0,-1.6) {a gene-product-to-\\term relation};
  \node[olog] (term) at (3.6,-1.6) {a GO term};
  \node[olog] (ns) at (1.8,-3.4) {a GO namespace\\(one of MF, BP, CC)};
  \draw[arr] (ann) -- node[above left,font=\scriptsize] {$g$} (gp);
  \draw[arr] (ann) -- node[left,font=\scriptsize] {$r$} (rel);
  \draw[arr] (ann) -- node[above right,font=\scriptsize] {$t$} (term);
  \draw[arr] (rel) -- node[left,font=\scriptsize] {applies to terms of\ } (ns);
  \draw[arr] (term) -- node[right,font=\scriptsize] {\ belongs to} (ns);
  \node at (1.8,-2.25) {$\checkmark$};
\end{tikzpicture}
\caption{Two Gene ologs. (a) Genes, transcripts and protein isoforms; the check mark declares the fact that the gene from which a transcript is transcribed is the gene that encodes its translation product. Arrows point towards \type{a gene} because a gene may have several transcripts and isoforms, or none. Hooked arrows (``is'') are injective. (b) GO annotations as a span (evidence omitted); the fact states that the namespace of the annotated term is the one to which the relation applies. Adapted from \citet{WuGeneOlogs}.}
\label{fig:olog}
\end{figure}

\begin{example}[Annotations as instance data]\label{ex:annotation}
A standard GO annotation relates a gene product $g$ to a GO term $t$ through a relation $r$ that depends on the namespace of $t$, such as ``enables'' (molecular function), ``involved in'' or ``acts upstream of or within'' (biological process) and ``located in'' (cellular component). This relationship is many-to-many, so it is a span (\cref{fig:olog}(b)). The check mark declares a fact with real content: any instance $I$ must satisfy
\[
I(\text{belongs to})\circ I(t) \;=\; I(\text{applies to terms of})\circ I(r),
\]
an integrity constraint that can be checked automatically on an annotation file. For three annotations of mouse Foxp1 the two sides agree:
\begin{center}
\small
\begin{tabular}{lllll}
\toprule
$g$ & $r$ & $t$ & via $t$ & via $r$\\
\midrule
Foxp1 & enables & DNA-binding transcription factor activity & MF & MF\\
Foxp1 & acts upstream of or within & regulation of action potential & BP & BP\\
Foxp1 & acts upstream of or within & innate vocalization behavior & BP & BP\\
\bottomrule
\end{tabular}
\end{center}
GO-CAM models are, in the same way, instance data on a fixed schema olog, in which activities, processes, locations and causal links are tables and their relations are columns or spans \citep{WuGeneOlogs}.
\end{example}

Universal constructions give GO terms a precise meaning. For instance, consider the term \type{a carbohydrate transmembrane transport}. Declaring the square formed by this term, its two parents \type{a carbohydrate transport} and \type{a transmembrane transport}, and their common parent \type{a transport} to be a \emph{pullback} asserts that every transport that is both a carbohydrate transport and a transmembrane transport is a carbohydrate transmembrane transport, which is the olog counterpart of an OWL definition by intersection. Functors between Gene ologs compare ontologies, and pushouts along a common sub-olog merge them, for example ologs built by different curation groups or for different species \citep{SpivakKent2012,WuGeneOlogs}.

\paragraph{Relation to OWL.} Gene ologs make functionality explicit, treat path equations as first-class constraints on data, define constructed types by universal properties, and place ontology, annotation data and their comparison in one framework. They are, like OWL, two-valued, and their computational cost at the scale of GO has not yet been studied. We therefore regard them as complementary to, not a replacement for, the OWL representation of GO \citep{WuGeneOlogs}.

\subsection{One pattern, three instances}\label{sec:pattern}

The three bridges look different, but they share one structure, the one described in \cref{sec:themes}: a category of syntax whose morphisms are built by composition, and a functor that assigns meaning compositionally (\cref{tab:pattern}). In each case the categorical formulation brings a concrete benefit: grammaticality and meaning of proteins with new domain combinations (language), construction of large networks and their dynamics from validated modules (networks), and checkable constraints and principled merging of knowledge (ontology).

\begin{table}[t]
\centering
\small
\begin{tabularx}{\textwidth}{>{\raggedright\arraybackslash}p{2.2cm} >{\raggedright\arraybackslash}X >{\raggedright\arraybackslash}X >{\raggedright\arraybackslash}X}
\toprule
 & Language (\cref{sec:language}) & Networks (\cref{sec:network}) & Ontology (\cref{sec:ontology})\\
\midrule
Syntax & pregroup reductions of domain architectures & open Petri nets $X\nrightarrow Y$ (cospans) & an olog (a finitely presented category)\\
Parts & domains and linkers & GRN modules & types, aspects, facts\\
Composition & concatenation $\otimes$ and contraction & gluing along shared genes (pushout); $\sqcup$ in parallel & composition of aspects; merging by pushout\\
Semantics functor & $F\colon\cat{Grammar}\to\cat{FVect}$ & operational, reachability (lax), rate equations & instances $I\colon\cat{C}\to\cat{Set}$\\
What is computed & a function vector in $P$ for every protein & processes, reachable states, dynamics of the whole network & a database of annotations satisfying the facts\\
\bottomrule
\end{tabularx}
\caption{The three bridges as instances of compositional, functorial semantics.}
\label{tab:pattern}
\end{table}

The bridges also have something else in common: they all concern genes and their products, but none of them describes the genome as such, that is, the arrangement of genes along chromosomes. We turn to this in the next section.

\section{A positional category of genes}\label{sec:genes}

We now take the second route and build a categorical structure for genomics from first principles. The most immediate structure of a genome is the arrangement of its genes along its chromosomes. A natural first candidate is therefore a category whose objects are the genes of a species and whose morphisms record their relative positions. Because relative position is a relation (one gene either lies before another or it does not), such a category is a preorder in the sense of \cref{ex:preorder}. The point of this section is to define it carefully, to show which choices are biological and which are conventions, and to show what it can be used for.

\subsection{Reference coordinates}\label{sec:coords}

\begin{figure}[t]
\centering
\includegraphics[width=0.62\textwidth]{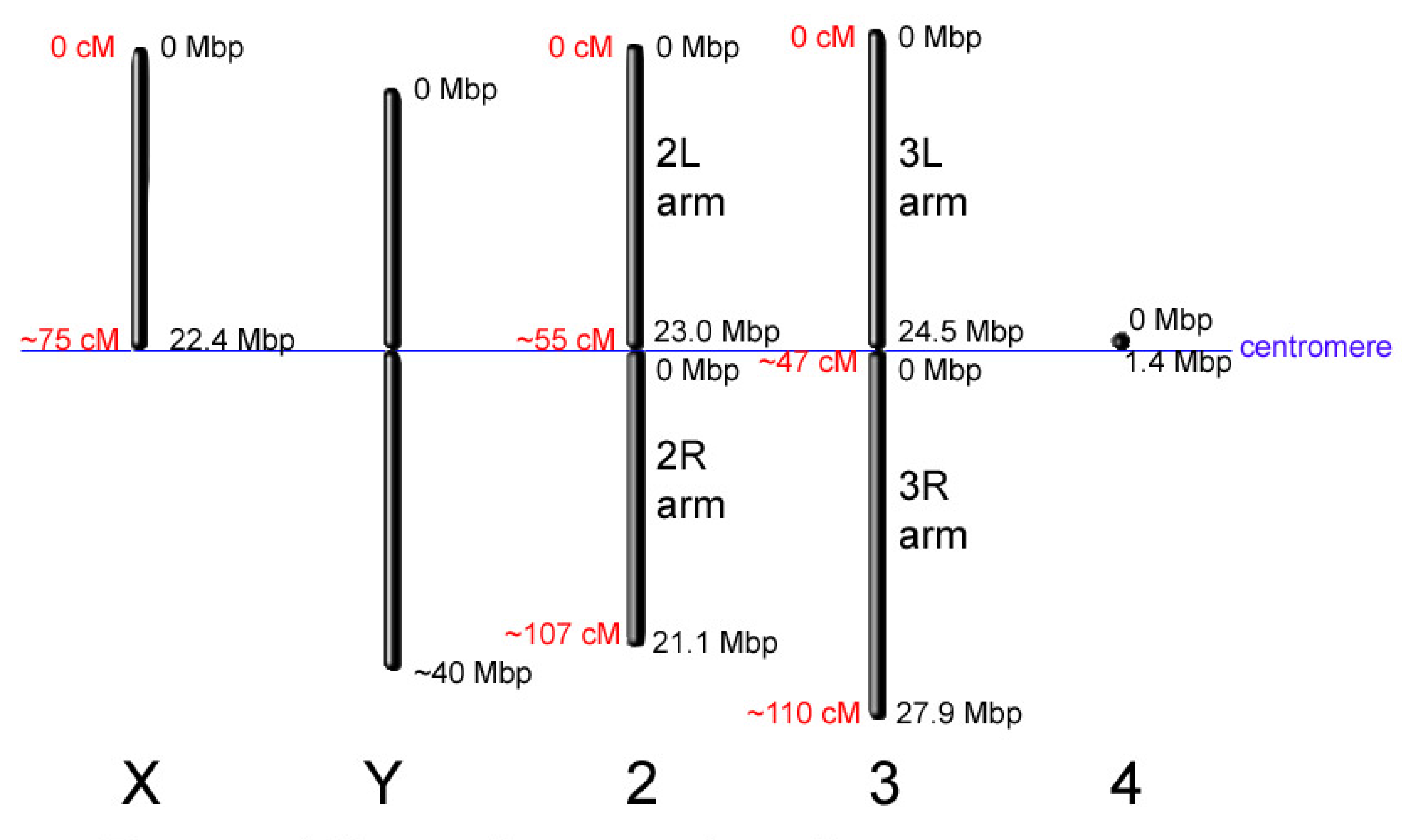}
\caption{\emph{Drosophila melanogaster} chromosomes, with approximate physical lengths (Mbp, black) and genetic map positions (cM, red). The arms of chromosomes 2 and 3 have separate coordinate systems, each starting from 0 (diagram by Steven J.\ Baskauf).}
\label{fig:chromosomes}
\end{figure}

Gene positions are given relative to a reference genome assembly. We use \emph{Drosophila melanogaster} as our example (\cref{fig:chromosomes}). A fly has four pairs of chromosomes: the sex chromosomes X and Y, and the autosomes 2, 3 and the small chromosome 4. Chromosomes 2 and 3 have their centromere near the middle, and their two arms are sequenced and annotated as separate sequences, 2L, 2R, 3L and 3R. Coordinates are counted along each arm separately: on 2L from the tip of the left arm towards the centromere, and on 2R from the centromere towards the tip of the right arm. The Release~6 reference assembly accordingly consists of the sequences X, 2L, 2R, 3L, 3R, 4 and Y, the mitochondrial genome, and a number of unplaced scaffolds \citep{Hoskins2015}. A ``position on chromosome 2'' is therefore not well defined without saying on which arm; and the 37 genes of the mitochondrial genome are part of the genome too.

A genome annotation (for example, a GFF file) records for each gene $g$ a reference sequence $\sigma(g)$, a \emph{start} $s(g)$ and an \emph{end} $e(g)$ with $1\le s(g)\le e(g)\le \ell_{\sigma(g)}$, where $\ell_\sigma$ is the length of $\sigma$, and a strand $\pm$. By convention $s(g)\le e(g)$ are coordinates on the forward strand \emph{whatever the strand of the gene}: a gene on the minus strand is transcribed from $e(g)$ towards $s(g)$. We assume that each gene has a single annotated locus; genes with several loci would be treated as several objects. Write $\mathcal G_\sigma$ for the set of genes on sequence $\sigma$.

\subsection{The interval order of genes}\label{sec:interval}

Two genes on the same sequence either overlap, or one of them lies entirely before the other. This gives the most basic positional relation.

\begin{definition}[Interval order of genes]\label{def:interval}
For $g,h\in\mathcal G_\sigma$, say that $g$ \emph{lies before} $h$, written $g\prec h$, if $e(g)<s(h)$. Write $g\sqsubseteq h$ if $g=h$ or $g\prec h$. The \emph{interval category} $\cat P_\sigma$ is the preorder $(\mathcal G_\sigma,\sqsubseteq)$ regarded as a category: there is one morphism $g\to h$ if $g\sqsubseteq h$ and none otherwise.
\end{definition}

\begin{proposition}\label{prop:interval}
$(\mathcal G_\sigma,\sqsubseteq)$ is a partial order. Two distinct genes are incomparable exactly when their intervals overlap.
\end{proposition}

\begin{proof}
Reflexivity holds by definition. If $g\prec h$ and $h\prec k$, then $e(g)<s(h)\le e(h)<s(k)$, so $g\prec k$. If $g\prec h$ and $h\prec g$, then $e(g)<s(h)\le e(h)<s(g)\le e(g)$, a contradiction; hence $\sqsubseteq$ is antisymmetric. Finally, two intervals $[s(g),e(g)]$ and $[s(h),e(h)]$ are disjoint exactly when $e(g)<s(h)$ or $e(h)<s(g)$.
\end{proof}

Partial orders of this kind are known as \emph{interval orders} \citep{Fishburn1985}. In $\cat P_\sigma$, overlapping genes, including nested genes and overlapping genes on opposite strands, are incomparable. This is biologically appropriate: such genes share sequence, and neither lies upstream of the other.

The interval order also captures the symmetry between the two strands. Reading the complementary strand in its own $5'\to3'$ direction reverses coordinates by the reflection $\rho(x)=\ell_\sigma+1-x$, which sends a gene's interval $[s(g),e(g)]$ to $[\rho(e(g)),\rho(s(g))]$.

\begin{proposition}[Strand symmetry]\label{prop:strand}
Let $\cat P^{\rho}_\sigma$ be the interval category of $\mathcal G_\sigma$ computed from the reflected coordinates $s^\rho(g)=\rho(e(g))$ and $e^\rho(g)=\rho(s(g))$. Then $\cat P^{\rho}_\sigma=\cat P_\sigma^{\op}$.
\end{proposition}

\begin{proof}
For distinct $g,h$, $g\prec^\rho h$ iff $e^\rho(g)<s^\rho(h)$ iff $\ell_\sigma+1-s(g)<\ell_\sigma+1-e(h)$ iff $e(h)<s(g)$ iff $h\prec g$.
\end{proof}

Thus the opposite category $\cat P_\sigma^{\op}$ is precisely the arrangement of genes as seen from the complementary strand. As we show next, this is true of the interval order but not of the linear order that is more commonly used.

\subsection{The linear order of genes, and the whole genome}\label{sec:linear}

For many purposes (sorting an annotation file, listing neighbouring genes) one wants a total ranking of the genes on a sequence. The ranking used in practice orders genes by start coordinate.

\begin{definition}[Linear order of genes]\label{def:linear}
For $g,h\in\mathcal G_\sigma$, let $g\le h$ if $(s(g),e(g),\mathrm{id}(g))\le(s(h),e(h),\mathrm{id}(h))$ in the lexicographic order, where $\mathrm{id}$ is a unique gene identifier. The resulting linear order, regarded as a category, is denoted $\cat L_\sigma$.
\end{definition}

In words: genes are ordered by their leftmost coordinate on the forward strand; genes with the same start are ordered with the shorter one first; and the identifier breaks any remaining tie. The first two rules are those of the first version of this paper. There, genes on the sense strand were ordered by start position and genes on the anti-sense strand by stop position; since a minus-strand gene stops at its leftmost coordinate $s(g)$, both rules amount to ordering by $s(g)$. Without a final tie-breaker, two genes with identical coordinates (for example on opposite strands) would satisfy $g\le h$ and $h\le g$, and $\le$ would only be a total preorder.

\begin{proposition}\label{prop:extension}
The identity on genes is a functor $\cat P_\sigma\to\cat L_\sigma$; that is, $\cat L_\sigma$ is a linear extension of the interval order $\cat P_\sigma$.
\end{proposition}

\begin{proof}
If $g\prec h$ then $s(g)\le e(g)<s(h)$, so $g<h$ in $\cat L_\sigma$.
\end{proof}

\begin{remark}[The opposite of the linear order]\label{rem:op}
Unlike $\cat P_\sigma$, the linear order is not strand-symmetric. Applying \cref{def:linear} to reflected coordinates orders genes by decreasing $e(g)$, whereas $\cat L_\sigma^{\op}$ orders them by decreasing $s(g)$. The two orders differ whenever one gene's interval contains another's: for $g=[100,500]$ and $h=[200,300]$ we have $g<h$ in $\cat L_\sigma$, so $h$ comes first in $\cat L_\sigma^{\op}$, whereas the reflected ranking puts $g$ first, because $g$ ends further to the right. Hence the opposite category of the gene order describes the order on the anti-sense strand only for the interval order (\cref{prop:strand}), or for linear orders of genes none of which is contained in another. This corrects the first version of this paper, which proposed the opposite of the linear order as a model of the anti-sense strand.
\end{remark}

\begin{figure}[t]
\centering
\begin{tikzpicture}[x=1cm,y=1cm]
  \node at (-0.4,2.3) {(a)};
  \node[font=\small, anchor=west] at (-0.4,1.7) {arm X};
  \draw[thick,gray] (1.2,1.7) -- (10.6,1.7);
  \fill[blue!60] (2.2,1.8) rectangle (3.2,1.98);
  \draw[-{Stealth[length=1.8mm]},blue!60!black] (2.2,2.15) -- (3.2,2.15);
  \node[font=\small,blue!60!black] at (3.45,1.9) {$A$};
  \node[font=\small, anchor=west] at (-0.4,0) {arm 2L};
  \draw[thick,gray] (1.2,0) -- (10.6,0);
  \node[font=\scriptsize,gray,anchor=west] at (10.65,0.12) {$+$};
  \node[font=\scriptsize,gray,anchor=west] at (10.65,-0.14) {$-$};
  \fill[orange!85] (1.776,0.42) rectangle (3.152,0.6);
  \fill[green!55!black] (1.776,0.12) rectangle (4.240,0.3);
  \fill[violet!80] (6.832,0.12) rectangle (9.312,0.3);
  \fill[olive!80!black] (6.000,-0.12) rectangle (7.360,-0.3);
  \node[font=\small] at (3.402,0.51) {$C$};
  \node[font=\small] at (4.490,0.21) {$B$};
  \node[font=\small] at (9.562,0.21) {$D$};
  \node[font=\small] at (7.610,-0.21) {$E$};
  \draw[-{Stealth[length=1.6mm]},olive!60!black] (7.360,-0.48) -- (6.000,-0.48);
  \draw[gray] (1.776,-0.75) -- (1.776,-0.85) node[below,font=\tiny] {256};
  \draw[gray] (3.152,-0.75) -- (3.152,-0.85) node[below,font=\tiny] {342};
  \draw[gray] (4.240,-0.75) -- (4.240,-0.85) node[below,font=\tiny] {410};
  \draw[gray] (6.000,-0.75) -- (6.000,-0.85) node[below,font=\tiny] {520};
  \draw[gray] (6.832,-0.75) -- (6.832,-0.85) node[below,font=\tiny] {572};
  \draw[gray] (7.360,-0.75) -- (7.360,-0.85) node[below,font=\tiny] {605};
  \draw[gray] (9.312,-0.75) -- (9.312,-0.85) node[below,font=\tiny] {727};
  \node at (-0.4,-1.7) {(b)};
  \node[font=\small] (hC) at (1.6,-3.3) {$C$};
  \node[font=\small] (hB) at (3.0,-3.3) {$B$};
  \node[font=\small] (hE) at (1.6,-2.1) {$E$};
  \node[font=\small] (hD) at (3.0,-2.1) {$D$};
  \draw[arr] (hC) -- (hE); \draw[arr] (hC) -- (hD);
  \draw[arr] (hB) -- (hE); \draw[arr] (hB) -- (hD);
  \node[font=\small] at (4.4,-2.7) {$A$};
  \node[font=\scriptsize] at (2.3,-3.8) {$\cat P_{\mathrm{2L}}$};
  \node[font=\scriptsize] at (4.4,-3.8) {$\cat P_{\mathrm X}$};
  \node at (5.6,-1.7) {(c)};
  \node[font=\small,anchor=west] at (6.0,-2.2) {$\cat L_{\mathrm{2L}}:\ C < B < E < D$};
  \node[font=\small,anchor=west] at (6.0,-2.8) {$\cat L_{\mathrm X}:\ A$};
  \node[font=\small,anchor=west] at (6.0,-3.4) {genome: $\cat G=\coprod_\sigma \cat L_\sigma$};
\end{tikzpicture}
\caption{Positional categories for five hypothetical genes, as in the first version of this paper. (a) Gene $A$ lies on arm X; genes $B$--$E$ lie on arm 2L, with forward-strand coordinates $C=[256,342]$, $B=[256,410]$, $E=[520,605]$ and $D=[572,727]$ (arbitrary units). Genes above the axis are on the plus strand; $E$ is on the minus strand and is transcribed leftwards. (b) The interval order $\cat P_{\mathrm{2L}}$ (arrows generate the morphisms): overlapping genes ($B$ and $C$; $D$ and $E$) are incomparable. Gene $A$ forms its own component. (c) The linear order $\cat L_{\mathrm{2L}}$ by start coordinate, with ties broken by length, extends $\cat P_{\mathrm{2L}}$ (\cref{prop:extension}). Genes on different sequences are not compared.}
\label{fig:genes}
\end{figure}

\paragraph{The whole genome.} The first version of this paper also ordered the chromosomes, as $\mathrm X\le\mathrm Y\le 2\le 3\le 4$, and placed every gene on an earlier chromosome before every gene on a later one. That ordering is a convention with no biological content; any other ordering of the sequences would serve equally well. The canonical construction is instead the \emph{coproduct} (disjoint union) of categories,
\[
\cat G=\coprod_{\sigma}\cat L_\sigma \qquad\text{and}\qquad \cat P=\coprod_\sigma \cat P_\sigma,
\]
in which genes on different sequences are simply not comparable. A single linear order on all genes, when it is wanted, is the \emph{ordinal sum} $\cat L_{\sigma_1}\oplus\cat L_{\sigma_2}\oplus\cdots$ for a chosen ordering $\sigma_1,\sigma_2,\dots$ of the sequences, and $\cat G$ includes into it by a functor. Some ordinal sums are meaningful: because coordinates on 2L run up to the centromere and those on 2R start from it, $\cat L_{\mathrm{2L}}\oplus\cat L_{\mathrm{2R}}$ is the order of genes along the whole of chromosome~2. \Cref{fig:genes} illustrates these constructions.

\subsection{What the positional category is for}\label{sec:using}

A category is useful through the functors into and out of it. We give three examples.

\paragraph{Coordinates.} The start coordinate is a functor $s\colon\cat L_\sigma\to(\N,\le)$, and the map sending a gene to its interval is a functor from $\cat P_\sigma$ to the interval order of all intervals of $\{1,\dots,\ell_\sigma\}$. Quantities defined along the genome, such as distances between neighbouring genes, are read off along these functors.

\paragraph{Synteny and inversions.} Comparing the genomes of two species or strains, orthology relates genes of one to genes of the other. A \emph{syntenic block}, a stretch of genes whose order is conserved, is a sub-order $B\subseteq\cat L_\sigma$ on which orthology is an order-preserving injection into some $\cat L'_\tau$, that is, a functor $B\to\cat L'_\tau$. A chromosomal \emph{inversion} reverses a segment, so on an inverted block orthology is order-reversing, a functor $B\to(\cat L'_\tau)^{\op}$. More precisely, an inversion reflects the coordinates of the segment, so by \cref{prop:strand} it sends the interval order of the segment exactly to its opposite, even when the segment contains overlapping genes. The arrangement of one genome relative to another is therefore described by a sequence of blocks, each mapped by a functor into either $\cat L'$ or $\cat L'^{\op}$. These are exactly the signed permutations studied in the theory of genome rearrangements, for example in sorting by reversals \citep{Hannenhalli1999}. Inversions are common in \emph{Drosophila}, both as natural polymorphisms \citep{Krimbas1992} and in the balancer chromosomes used in fly genetics. Here the opposite category has a genuine biological role.

\paragraph{Nested chromatin domains.} TADs and their sub-domains are also intervals on the reference coordinates, so the interval order applies to them as well. Domains at one level of the hierarchy that do not overlap form a linear order, and containment of sub-domains in domains forms a forest, a kind of partial order \citep[Ex.~4.4.2.5]{Spivak2014}. Sending each gene to the domain of a given level that contains it is a functor from (the genes lying within domains of) $\cat P_\sigma$ to the order of domains: if $g\prec h$ with $g$ in domain $T_1$ and $h$ in domain $T_2\neq T_1$, then $T_2\prec T_1$ would give $e(h)\le e(T_2)<s(T_1)\le s(g)\le e(g)<s(h)\le e(h)$, which is impossible, so $T_1\prec T_2$. Coarsening from sub-domains to domains is likewise a functor, so the hierarchy of the 3D genome appears as a chain of functors
\[
\cat P_\sigma \longrightarrow \cat D^{(k)}_\sigma \longrightarrow \cat D^{(k-1)}_\sigma \longrightarrow\cdots\longrightarrow\cat D^{(1)}_\sigma,
\]
from genes to the finest domains and on to the coarsest. This is a first, purely positional, step towards describing the hierarchical feature of genomic data (\cref{sec:next}) compositionally.

The positional category is deliberately simple. It records where genes are, not what they do. Its value lies in serving as a common reference to which the functional views of \cref{sec:bridges} can be attached, as we now discuss.

\section{Towards integration}\label{sec:integration}

Each of the four views developed above concerns genes, or the products of genes: as words assembled into proteins (\cref{sec:language}), as places in regulatory networks (\cref{sec:network}), as annotated entities in an ontology (\cref{sec:ontology}) and as positioned objects in a genome (\cref{sec:genes}). Genes are therefore the natural point at which to connect them. Ologs, which are at once knowledge representations and database schemas, provide the means.

\Cref{fig:schema} sketches a schema olog connecting the views. Every arrow is functional under the stated modelling assumptions (for example, a single annotated locus per gene), and many-to-many relationships such as regulation and annotation are spans. An instance $I$ of this schema is a database. Each view of the previous sections is computed from it.
\begin{itemize}
\item \emph{Position.} The tables \type{a gene} and \type{an interval on a reference sequence} give $\cat P_\sigma$ and $\cat L_\sigma$ (\cref{sec:genes}); the same construction applied to \type{a chromatin domain} gives the domain orders and the functors between them (\cref{sec:using}).
\item \emph{Language.} The domain occurrences of a protein isoform, ordered by their positions along the protein, give its domain architecture. This is the interval-order construction of \cref{sec:interval} again, now at the scale of a protein. Together with domain vectors, it provides the input to the DisCoCat model (\cref{sec:language}).
\item \emph{Networks.} The span \type{a regulatory interaction} gives the places and transitions of a Petri net (\cref{sec:network}), and any partition of the interactions into modules gives open Petri nets whose interfaces are computed from the shared genes, following rules (R1)--(R3).
\item \emph{Ontology.} The span \type{an annotation} is the annotation instance of \cref{ex:annotation}, and the GO terms connect to a Gene olog of GO itself.
\end{itemize}
Data from different sources (a genome annotation, a regulatory database, a domain database, GO annotation files) usually come with their own schemas. A functor from such a source schema to the common schema induces data migration functors that move the data in a canonical way \citep{Spivak2012}, \citep[Sec.~7.1.4]{Spivak2014}. This is the precise sense in which category theory can turn the integration of heterogeneous genomic data into a mathematical construction rather than an ad hoc one.

\begin{figure}[t]
\centering
\begin{tikzpicture}[x=1cm,y=1cm, every node/.style={font=\small}]
  \node[olog] (term) at (-5.2,4.4) {a GO term};
  \node[olog] (dfam) at (0,4.4) {a domain family};
  \node[olog] (docc) at (5.2,4.4) {a domain occurrence};
  \node[olog] (ann) at (-5.2,2.2) {an annotation};
  \node[olog] (gp) at (0,2.2) {a gene product};
  \node[olog] (iso) at (5.2,2.2) {a protein isoform};
  \node[olog] (reg) at (-5.2,0) {a regulatory interaction};
  \node[olog] (gene) at (0,0) {a gene};
  \node[olog] (dom) at (-5.2,-2.2) {a chromatin domain};
  \node[olog] (int) at (0,-2.2) {an interval on a\\ reference sequence};
  \node[olog] (strand) at (5.2,-2.2) {a strand ($+$ or $-$)};
  \draw[arr] (ann) -- node[left,font=\scriptsize] {has as term} (term);
  \draw[arr] (ann) -- node[above,font=\scriptsize] {has as subject} (gp);
  \draw[arr] (docc) -- node[above,font=\scriptsize] {belongs to} (dfam);
  \draw[arr] (docc) -- node[right,font=\scriptsize] {occurs in} (iso);
  \draw[arr,{Hooks[left]}-{Stealth[length=2.2mm]}] (iso) -- node[above,font=\scriptsize] {is a} (gp);
  \draw[arr] (gp) -- node[right,font=\scriptsize] {is encoded by} (gene);
  \draw[arr] ([yshift=2pt]reg.east) -- node[above,font=\scriptsize] {has as regulator} ([yshift=2pt]gene.west);
  \draw[arr] ([yshift=-2pt]reg.east) -- node[below,font=\scriptsize] {has as target} ([yshift=-2pt]gene.west);
  \draw[arr] (gene) -- node[right,font=\scriptsize] {occupies} (int);
  \draw[arr] (dom) -- node[above,font=\scriptsize] {occupies} (int);
  \draw[arr] (gene) -- node[above right,font=\scriptsize] {is on} (strand);
\end{tikzpicture}
\caption{A schema olog connecting the four views of genes. Spans (\type{a regulatory interaction}, \type{an annotation}) represent many-to-many relationships. An instance of this schema is a database from which the positional categories of \cref{sec:genes}, the domain architectures used by the DisCoCat model, the Petri nets of regulatory networks and GO annotation data are all computed. Further aspects (the sign of a regulatory interaction, the relation and evidence of an annotation, the position of a domain occurrence within its isoform) are omitted for clarity.}
\label{fig:schema}
\end{figure}

The integrated view suggests questions that no single view can ask. For example, the functor from genes to chromatin domains (\cref{sec:using}) partitions a regulatory network into modules, one for each domain, whose interfaces consist of the genes regulated across domain boundaries. Rules (R1)--(R3) guarantee that these modules recompose into the whole network. One can then ask whether such 3D-defined modules have simpler interfaces, or more nearly one-way interactions, than modules chosen at random, a question that combines the positional, network and 3D views precisely.

\paragraph{Open problems.} We close this section with directions in which categorical genomics can be developed.
\begin{enumerate}
\item \emph{Protein language.} Test the distributional and compositional hypotheses of the DisCoCat model on proteome-scale data, and develop a grammar of interactions between proteins, the analogue of DisCoCirc, to go from proteins to complexes and pathways.
\item \emph{Network semantics.} Equip open Petri net models of GRNs with further semantics (stochastic, Boolean, other rate laws) and develop compositional reachability analysis for GRN modules.
\item \emph{Ontology at scale.} Build Gene ologs at the scale of GO, address negation and uncertainty, and compare their computational cost with OWL reasoning.
\item \emph{Beyond the reference genome.} Extend the positional category from a single reference to pangenomes \citep{Liao2023}, in which the genomes of many individuals are represented together, and from the linear order to the 3D genome, in which chromatin contacts are a weighted relation between loci.
\item \emph{Software.} Implement these constructions with existing ACT software, such as AlgebraicJulia \citep{Patterson2022} and \texttt{lambeq} \citep{Kartsaklis2021}, so that they can be applied to real genomic data.
\end{enumerate}

\section{Conclusion}\label{sec:conclusion}

Genomics faces the challenge of integrating heterogeneous, hierarchical and densely interconnected data, and applied category theory offers a language built for composition and for structure-preserving comparison. We have argued that the two fields are a natural match and given evidence on two fronts. First, three existing bridges, through language, networks and ontology, are all instances of one pattern, compositional and functorial semantics, and each delivers a concrete benefit for genomics. Second, even the simplest structure of a genome, the arrangement of its genes, gives a category with useful properties: an interval order whose opposite is the view from the other strand, a linear order that extends it, and functors that describe synteny, inversions and nested chromatin domains. These views can be connected through a common schema.

These results are first steps. By naming the programme \emph{categorical genomics} and laying out its motivation, tools and open problems, we hope to draw the attention of both category theorists and genomicists, and to encourage them to work together on the organisational principles of the genome.

\section*{Acknowledgements}

The author thanks Quanlong Wang for introducing category theory to her, and Richard Southwell, Eugenia Cheng, John Baez, Bob Coecke, David Spivak, Brendan Fong, Bartosz Milewski, Emily Riehl, Steve Awodey, Tai-Danae Bradley and many others for their efforts in making category theory accessible to an audience beyond mathematicians. The author also thanks Claude Opus 5.5 (Anthropic) for assistance in checking and revising the manuscript and in preparing the figures; the author takes full responsibility for the content.

\bibliographystyle{plainnat}
\bibliography{refs}

\end{document}